\documentclass[prl,twocolumn,showpacs,superscriptaddress,floatfix]{revtex4}
\usepackage{graphicx}
\usepackage{amssymb}
\usepackage{subfigure}

\newcommand{\bea}{\begin{eqnarray}}
\newcommand{\eea}{\end{eqnarray}}
\newcommand{\bc}{\begin{center}}
\newcommand{\ec}{\end{center}}
\newcommand{\beqa}{\begin{eqnarray}}
\newcommand{\eeqa}{\end{eqnarray}}
\newcommand{\bay}{\begin{array}}
\newcommand{\eay}{\end{array}}
\newcommand{\btab}{\begin{tabular}}
\newcommand{\etab}{\end{tabular}}

\newcommand{\szz}{\sigma_{zz}}

\newcommand{\sab}{\sigma_{\alpha\beta}}

\renewcommand{\vec}[1]{{\bf #1}}

\begin{document}

\title{
Scale separation in granular packings: stress plateaus and fluctuations
}

\author{C. Goldenberg}
\email{chayg@lpmcn.univ-lyon1.fr}
 \altaffiliation{Present address:
Laboratoire de Physique de la Mati{\`e}re Condens{\'e}e et Nanostructures,
Universit{\'e} Lyon 1; CNRS, UMR 5586, Domaine Scientifique de la Doua, F-69622
Villeurbanne cedex; France.}
\affiliation{Laboratoire de Physique et M\'ecanique des Milieux
  H\'et\'erog\`enes (CNRS UMR 7636), ESPCI, 10 rue Vauquelin, 75231 Paris Cedex
  05, France.}
\author{A.~P.~F. Atman}
\email{atman@fisica.ufmg.br}
\affiliation{Departamento de F{\'i}sica, Instituto de Ci{\^e}ncias Exatas,
Universidade Federal de Minas Gerais, C.P. 702, 30123-970, Belo Horizonte, MG,
Brazil.}
\author{P. Claudin}
\email{claudin@ccr.jussieu.fr}
\affiliation{Laboratoire de Physique et M\'ecanique des Milieux
  H\'et\'erog\`enes (CNRS UMR 7636), ESPCI, 10 rue Vauquelin, 75231 Paris
  Cedex 05, France.} 
\author{G. Combe}
\email{gael.combe@ujf-grenoble.fr}
\affiliation{Laboratoire Interdisciplinaire de Recherche Impliquant la G\'eologie et
la M\'ecanique,\\BP 53, 38041 Grenoble Cedex 09, France.}
\author{I. Goldhirsch}
\email{isaac@eng.tau.ac.il}
\affiliation{Department of Fluid Mechanics and Heat Transfer,
Faculty of Engineering, Tel Aviv University,
Ramat-Aviv, Tel Aviv 69978, Israel.}

\date{\today}

\begin{abstract}
  It is demonstrated, by numerical simulations of a 2D assembly of polydisperse
  disks, that there exists a range (plateau) of coarse graining scales for
  which the stress tensor field in a granular solid is nearly resolution
  independent, thereby enabling an `objective' definition of this field.
  Expectedly, it is not the mere size of the the system but the (related)
  magnitudes of the gradients that determine the widths of the plateaus.
  Ensemble averaging (even over `small' ensembles) extends the widths of the
  plateaus to sub-particle scales.  The fluctuations within the ensemble are
  studied as well.  Both the response to homogeneous forcing and to an external
  compressive localized load (and gravity) are studied.  Implications to
  small solid systems and constitutive relations are briefly discussed.
\end{abstract}

\maketitle

Continuum descriptions of matter comprise equations of motion for appropriate
sets of macroscopic fields~\cite{Truesdell66}. It is helpful (though not
required) when these fields or the constitutive relations do not depend on the
averaging scale within a certain range of scales (larger than microscopic and
smaller than the scales characterizing field gradients), see e.g.,
\cite{Batchelor67}.  These `plateaus' of scales define  the
macroscopic fields and their fluxes in a ``scale independent'' way. The
corresponding constitutive relations are local (expressed in terms of gradients
of the fields) when scale separation exists.  The Navier-Stokes equations  and solid elasticity are typical examples.  

Recent studies render support to the notion of scale separation in nanoscale
solids and granular matter, as their mechanics lends itself to description by
elasticity for a range of loads
\cite{Reydellet01,Serero01,Wittmer02,Goldhirsch02,Leonforte04,Goldenberg05,
  Atman05,Leonforte05}.  The short correlation length of the contact forces in
static granular matter \cite{ForceCorrelations} suggests that such plateaus
should indeed exist.   In this Letter, we study the scale
dependence of the stress response in polydisperse granular packings. We show
that even in small systems where stress gradients may be large, stress plateaus
can be identified (though they may be quite narrow in some cases); expectedly,
the widths of the plateaus are related to the gradients of the stress field.
Ensemble averaging increases these widths. The scale dependence of the
fluctuations within the ensemble is studied as well.

The considered model is a two-dimensional (2D) rectangular slab in the $x-z$
plane, periodic in the $x$ direction. It comprises $3600$ polydisperse
frictional disks, whose radii are uniformly distributed in the range $\left(
  R_{\rm min}, R_{\rm max} = 2 R_{\rm min}\right)$. The aspect ratio is about
7.  Gravity acts in the $-\hat{{\bf z}}$ direction.  It is prepared by
sequentially dropping the grains from rest at random horizontal positions above
the top of the system.  The floor comprises about $160$ non-touching grains,
whose centers are constrained to reside at $z=0$, and whose radii are randomly
chosen from the range $\left(1.2 R_{\rm min}, R_{\rm max}\right)$, which
ensures that bulk particles cannot percolate through the floor.  Similar
systems have been studied experimentally both in two and three dimensions
\cite{Reydellet01,Serero01,Mueggenburg02,Atman05b}.

The force model used in the simulations is essentially that of Cundall and
Strack~\cite{Cundall79}: overlapping disks are coupled by both normal and
tangential springs, of respective stiffnesses, $k_n=1.5\cdot 10^4
\frac{\left<m\right>g}{R_{{\mbox{\tiny{max}}}}}$ and $k_t = 0.5 k_n$, where
$\left<m\right>$ is the mean particle mass and $g$ the gravitational
acceleration. A linear viscous damping force (dashpot) acts in the normal
direction (parallel to the line connecting the centers of the disks), the
damping coefficient chosen to correspond to critical damping.  The tangential
forces are limited by the Coulomb condition, with a coefficient of (both static
and dynamic) friction, $\mu = 0.5$.  The simulation is run until it reaches a
numerically static state~\cite{AtmanIP}. To this system we apply either a
uniform external load (at its top) or a vertical compressive force acting on
one particle at the top of the system.  In the latter case the displacement
gradients are large near the point of application of the force and decay with
distance from this point; the effects of the local gradients on the widths of
the plateaus can thus be studied in the same system.  The external load is
linearly increased from zero to its final value, $F_0$, in a time comparable to
the typical relaxation time to static equilibrium.  The load is subsequently
kept fixed, and the system is relaxed to a new numerically static state.  The
system's response is linear in the magnitude of the force for loads not
exceeding a few times $\left<m\right>g$, as in Ref.~\cite{Goldenberg05}. The
response obeys superposition when two external forces are applied, and is
reversible to the slow removal of the force~\cite{AtmanIP}. Below we specialize
to the range of loads for which the response is linear. Findings for single
realizations as well as `ensemble averages' over different realizations of the
disorder are presented.  The results reported here pertain to the normal stress
{\em response} of the system (at the floor), i.e., the difference between the
floor stress for the loaded system and the (near-uniform) floor stress of the
unloaded system.

The stress field, $\sigma_{\alpha\beta}({\bf r})$, at point ${\bf r}$, is given
(without the kinetic stress term, which vanishes in the static limit) by the
following exact expression, which is fully compatible with the general
equations of continuum mechanics (for both static and time dependent states)
\cite{Glasser01,Goldhirsch02,Babic97,Zhu02}:
\begin{equation}
\label{eq:defsab}
\sab(\vec{r}) = \frac{1}{2} \sum_{i,j; i \neq j} f_{ij \alpha} r_{ij \beta}
\int^1_0 \!\! ds \, \phi[\vec{r} - \vec{r}_i + s \vec{r}_{ij}],  
\end{equation} 
where $i,j$ are particle labels, $\alpha,\beta$ represent Cartesian components,
$\vec{r}_{ij} \equiv \vec{r}_i - \vec{r}_j$, where $\vec{r}_i$ is the center of
mass of particle $i$, and $\vec{f}_{ij}$ is the force exerted by particle $j$
on particle $i$. The coarse graining (CG) function, $\phi(\vec{R})$, is a
positive semidefinite normalized function, with a single maximum at ${\bf
  R}=0$, and width $w$ (the CG scale).  The sign convention here is that
compressive stress is positive.

The validity of the
equations of continuum mechanics (unlike the constitutive relations) is not
resolution limited [with Eq.~(\ref{eq:defsab}) defining the stress field],
see e.g., \cite{Glasser01,Goldhirsch02}.  The
actual values of the stress depend on the choice of $\phi$ and $w$.  For
`large' CG scales the dependence of the stress on the choice of $\phi$ is weak
for `reasonable'  CG functions, such as  
\cite{Glasser01,Goldenberg02}:  $\phi(\vec{R})= 1/(\pi w^2)
H(w-|\vec{R}|)$, where $H$ is the Heaviside function, and the Gaussian,
$\phi(\mathbf{r})=\frac{1}{\pi w^{2}}e^{-(|\mathbf{r}|/w)^{2}}$ (in 2D). 
  Gaussian CG functions yield smoother stress
fields than the Heaviside function, for obvious reasons~\cite{AtmanIP}. The
stress field of Eq.~(\ref{eq:defsab}) corresponds to the
Born-Huang~\cite{Born88} (or Irving-Kirkwood~\cite{Irving50}) formula in the
limit of large CG scales.

Experimentally~\cite{Reydellet01,Serero01}, the normal stress at the floor,
$\szz$, is given by: $\szz (x) = - \frac{1}{L} \sum_i f_{iz} H(\frac{L}{2}-
|x-x_i|)$, where $L$ is the CG length (gauge area in experiments), $f_{iz}$ is
the force acting on the floor particle $i$ and $x_i$ is its position.  An
identical expression can be obtained from Eq.~(\ref{eq:defsab}) by substituting
the anisotropic CG function $\phi(\vec{r})= \frac{1}{L} H(\frac{L}{2}- |x-x_i|)
\delta(z)$.

As a theoretical basis for defining  ensembles for 
 granular solids is lacking, one usually averages (with equal weights) the desired
entities over randomly generated realizations, subject to some constraints,
such as the construction protocol of the system. It is not a-priori
clear~\cite{Goldenberg04b} whether these
averages indeed yield typical values.  In order to study the effect of ensemble
averaging, we prepared $10$ different samples; this rather small ensemble was
extended as follows.  To each of the realizations we applied an external load
$F_0=7\left<m\right>g$ to different particles at the top of the slab.  We
verified that the corresponding fluctuations were nearly statistically
independent when these positions were separated by distances exceeding about
half the height of the system.  The CG response profiles were therefore
ensemble averaged over ensembles of effective sizes, $N_e$, up to $N_e=110$.
The $x$-coordinate of the load defines $x=0$.
\begin{figure}[b]
\bc
\includegraphics[clip,width=\hsize,clip,angle=0]{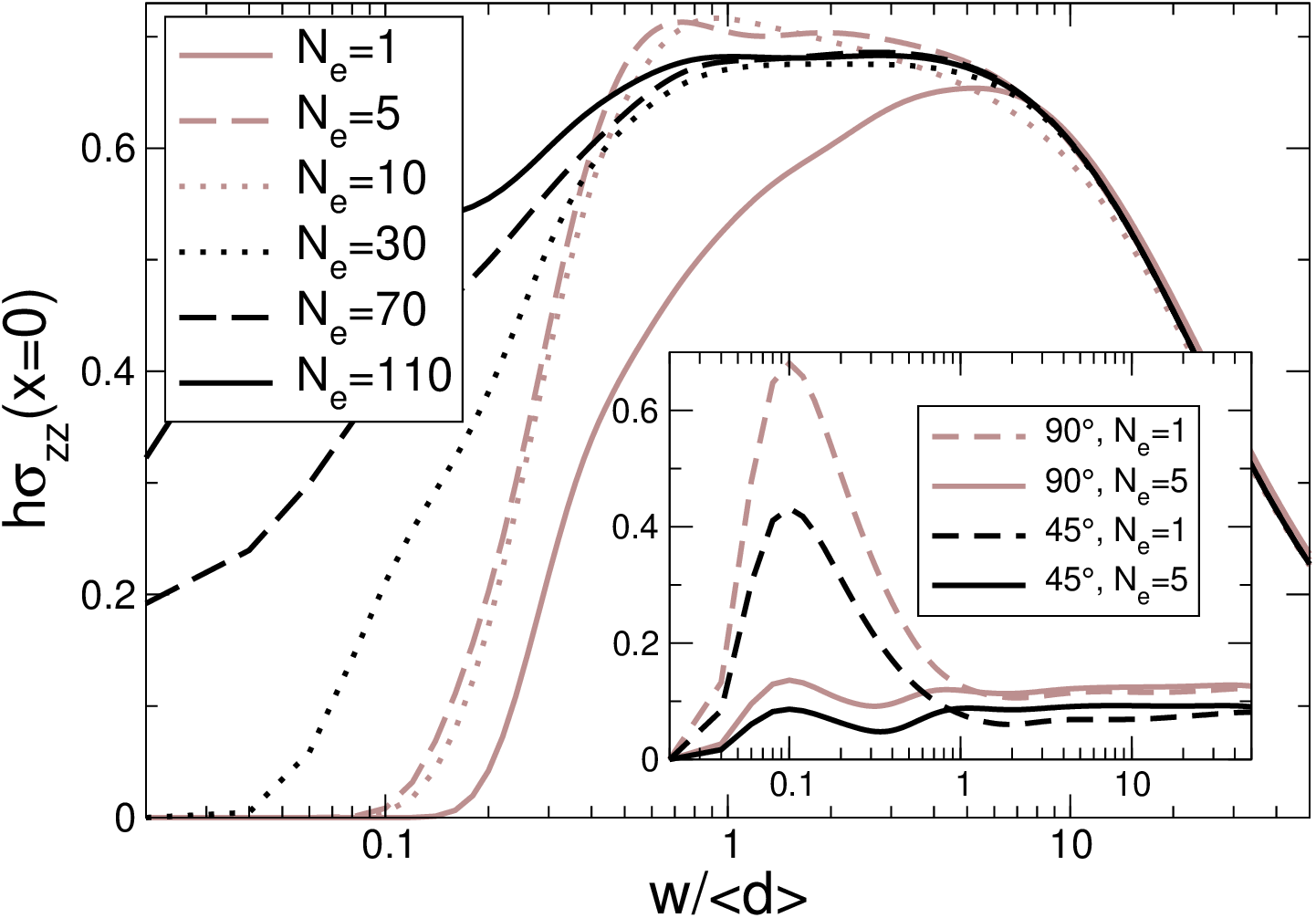}
\ec
\caption{Mean response to a   force at $x=0$,  vs.\ CG width,  $w$,
for different   ensemble sizes, $N_e$. The CG function is Gaussian,
and the unit of  $\sigma_{zz}$ is   $F_0/l$,  where
 \mbox{$l\approx 160\left<d\right>$} is the
 width of the system; $h$ is the height of the system. 
Inset: the response to homogeneous forcing at
  $45^\circ$ and $90^\circ$ to the horizontal.
  \label{fig:sx0_vs_w}}
\end{figure}

The inset of Fig.~\ref{fig:sx0_vs_w} presents the response of the system to
uniform forcing at its top both for the case of vertical forces (uniaxial
vertical stress) and oblique ones (a combination of vertical stress and
horizontal shear).  Forces of magnitude $7\left<m\right>g/n_t$, were applied to
each of the $n_t \sim 150$ particles whose centers resided at $z\geq
20.5\left<d\right>$, where $\left< d \right>$ denotes the average diameter of a
particle.  The response is quite flat for $w \geq \left< d \right> $ even for a
single realization, and flatter for an average over five realizations. In
contrast, in the inhomogeneous case (with a localized force at $x=0$), we
obtain narrower plateaus. Fig.~\ref{fig:sx0_vs_w} depicts the response at $x=0$
in this case.  Even for a single realization, one observes a (narrow) plateau.
The deviations from the plateau at large values of $w$ are due to the large
scale (macroscopic) spatial dependence of the average stress.  For the case
studied here, the mean response function (for $w$ of a few particle diameters
and $N_e=110$; see Fig.~\ref{fig:smeansdevs_vs_absx_diffw}) can be well fitted
by a Gaussian of half-width,  $W\simeq 17.5\left<d\right>$ (except at the
tail), hence (ignoring small scale fluctuations) the convolution involved in
the coarse graining process yields, for $w >> d$, a (approximate) Gaussian of
half-width $\sqrt{W^2+w^2}$, in agreement with the results obtained by direct
coarse-graining.

\begin{figure}[t]
\bc
\includegraphics[clip,width=\hsize,clip,angle=0]{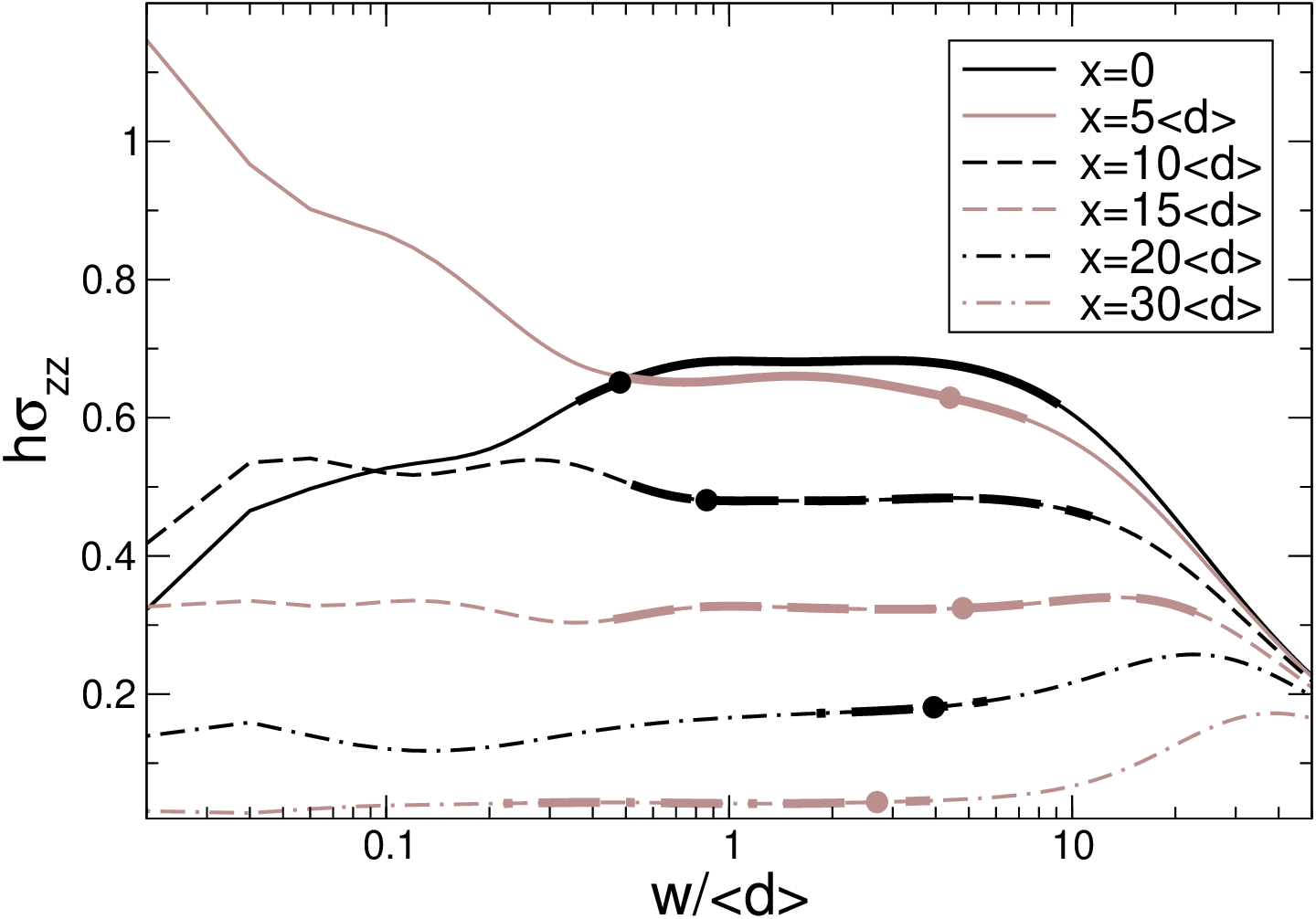}
\ec
\caption{Mean response at different $x$-coordinates at the floor
vs.\ CG width, $w$, for an ensemble size $N_e=110$. The thickened parts 
 correspond to the plateaus;  $\bullet$ marks  $w_0$. 
  \label{fig:sdiffx_vs_w}} \end{figure}

\begin{figure}[t]
  \bc
\includegraphics[clip,width=\hsize,clip,angle=0]{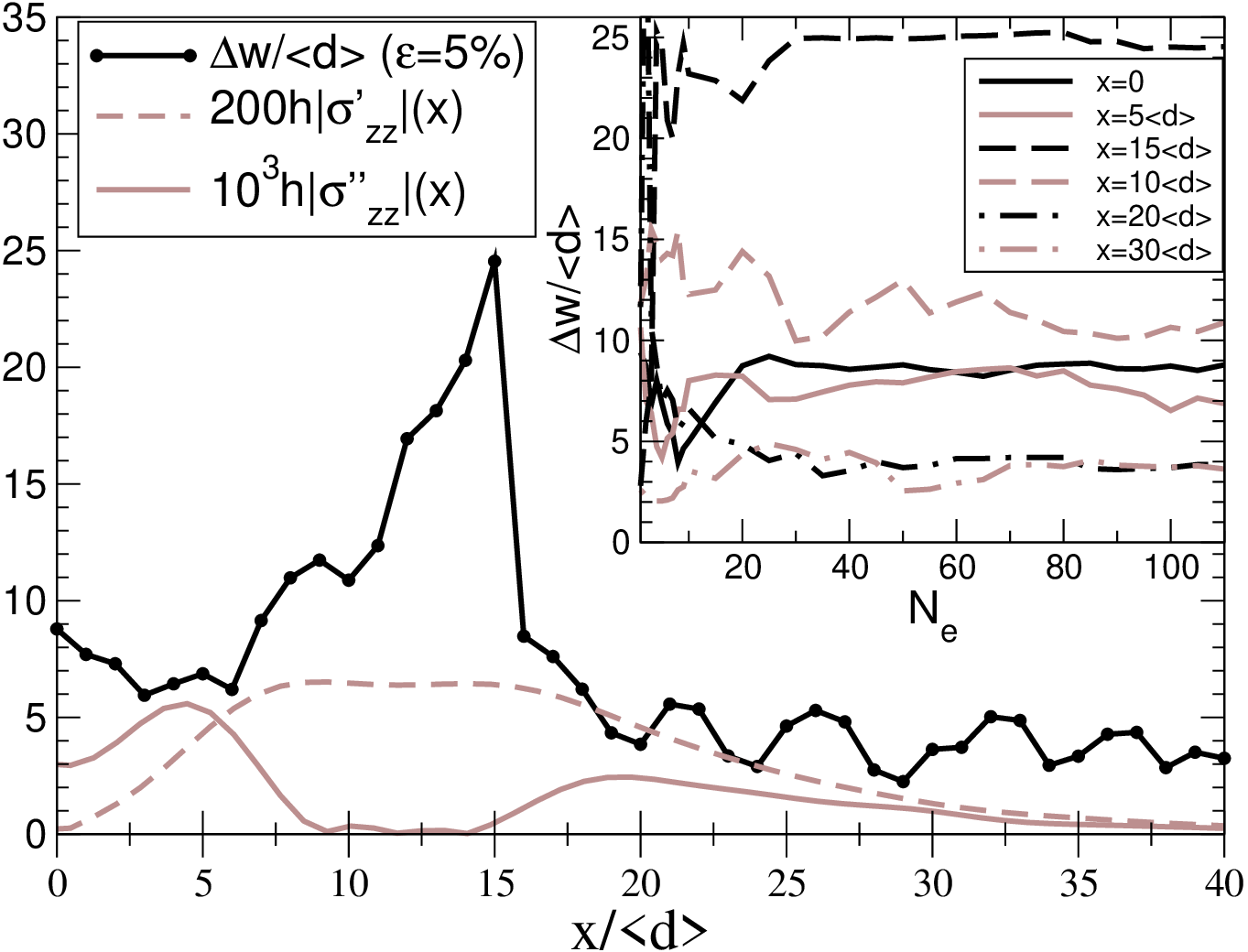}
\ec
\caption{The width of the plateau, $\Delta w$, vs.\ the $x$-coordinate, for an
  ensemble size $N_e=110$ and tolerance $\epsilon=5\%$, as well as the
  first and second derivatives of the response (for $w=4\left<d\right>$).
  Inset: $\Delta w$ vs.\ $N_e$ at different $x$-coordinates.
  \label{fig:deltaw}}
\end{figure}

\begin{figure}[h!]
\bc
\includegraphics[clip,width=\hsize,clip,angle=0]{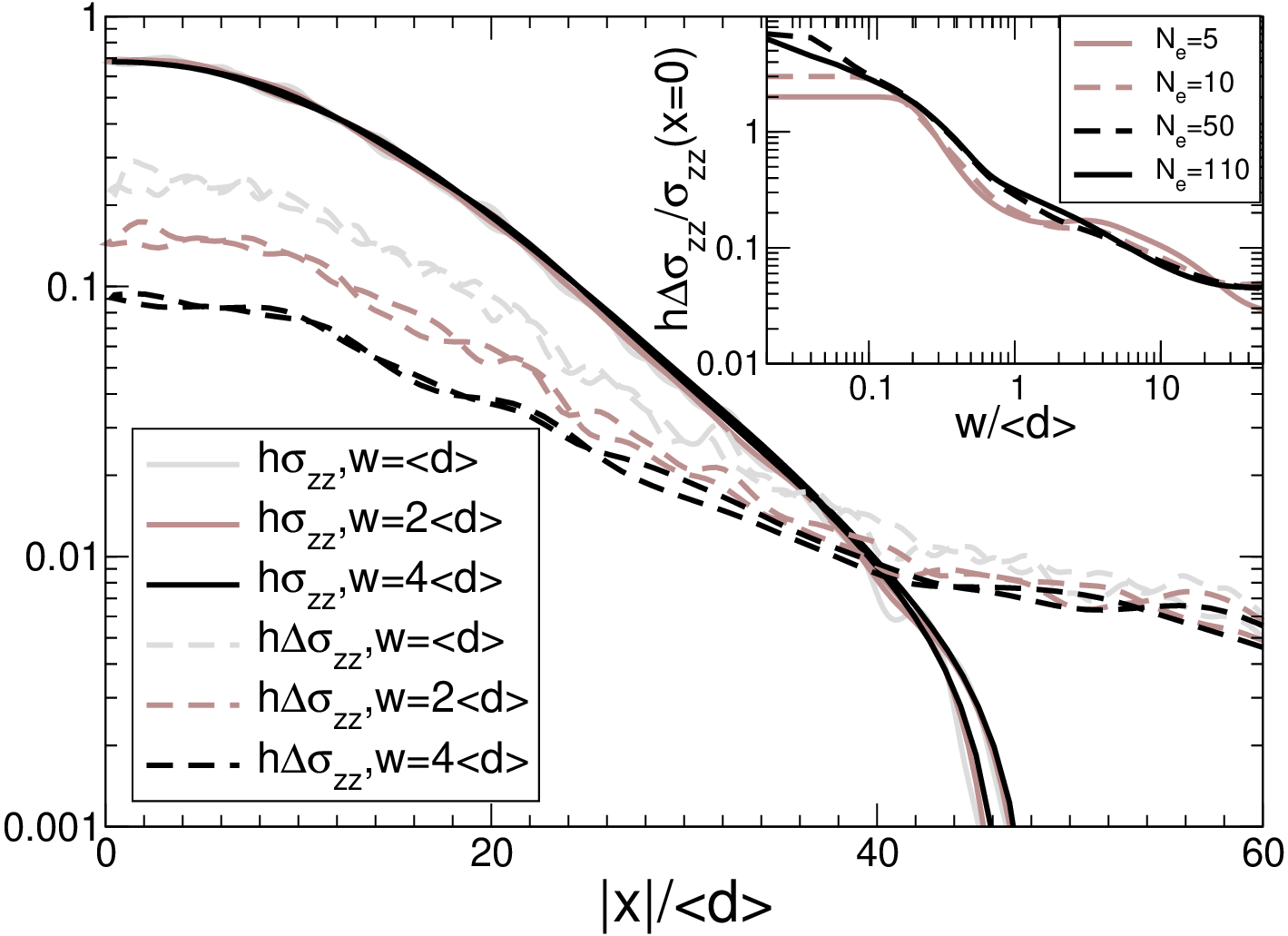}
\ec
\caption{Mean response and its standard deviation vs.\ $|x|$ at the floor, for
  several values of the CG width $w$, for an ensemble size $N_e=110$. Inset:
  Relative standard deviation of the response at $x=0$, vs.\ $w$, for
  different $N_e$.
  \label{fig:smeansdevs_vs_absx_diffw}}
\end{figure}

The presence of a plateau suggests that the stress is `locally homogeneous';
however, since the forcing is macroscopically inhomogeneous, the width of the
plateau depends on position, as shown in Fig.~\ref{fig:sdiffx_vs_w}. In order
to study this dependence in more detail, we define the plateau width $\Delta w$
as the largest connected range of $w$ for which $\left| \left[
  \sigma_{zz}(w)/\sigma_{zz} (w_0)\right] -1 \right|<\epsilon$, for a given tolerance
$\epsilon$ (maximizing $\Delta w$ over $0<w_0<5 \left<d\right>$).
Fig.~\ref{fig:deltaw} presents the dependence of $\Delta w$ on $x$ for
$\epsilon=5\%$ and $N_e=110$. The plateau widths can be rather small (a few
particle diameters) in parts of the system, as may be expected considering the
inhomogeneity of the response in this relatively small system.  The dependence
of $\Delta w$ on $x$ is related to the shape of the macroscopic response.  One
expects $\Delta w$ to be smaller where the response is less homogeneous. In
Fig.~\ref{fig:deltaw} we plot the first and second derivatives of the response
(calculated for $w=4\left<d\right>$). In general, $\Delta w$ appears to be
anti-correlated with the second derivative of the response, rather than the
first derivative. The reason seems to be that the average over a symmetric
segment around $x$, of a profile which is locally linear (on average) and whose
fluctuations are essentially uncorrelated, yields basically the same
(coarse-grained) value at $x$ almost irrespective of the width of the segment,
$w$. It is easy to verify that the size of the plateau should be given by
$(\Delta w)^2 \left| \frac{\partial ^2 \sigma_{zz}}{\partial x^2}\right|
\approx \epsilon \left|\sigma_{zz}\right|$; the peak in Fig.~\ref{fig:deltaw}
corresponds to a region where $\frac{\partial ^2 \sigma_{zz}}{\partial x^2}$
vanishes and there the plateau size depends on the next even (fourth)
derivative of $\sigma_{zz}$.

The dependence of $\Delta w$ on $N_e$ for several values of $x$ is shown in the
inset of Fig.~\ref{fig:deltaw}. As expected, ensemble averaging smoothes small
scale fluctuations: while for $N_e=1$ the plateaus start at $w\gtrsim
\left<d\right>$ (see also Fig.~\ref{fig:sx0_vs_w}), they probably extend down
to $w=0$ as $N_e\to \infty$.  Note that not only the forces fluctuate among the
realizations, but also the particle positions.  The plateau widths seem to
practically saturate already for $N_e \approx 50$, suggesting that the
fluctuations in the ensemble are not strong.

The wide distribution of contact {\em forces} in granular
materials~\cite{ForceCorrelations,Blair01} may suggest the existence of strong
stress fluctuations.  However, since the force correlations are rather short
ranged~\cite{ForceCorrelations}, their fluctuations are well smoothed by
spatial averaging. Denote the standard deviation (in the ensemble) of the
response by $\Delta \sigma$.  The relative standard deviation, $\Delta
\sigma_{zz}/\sigma_{zz}$, as a function of the CG scale $w$, for different
ensemble sizes, is presented in the inset of
Fig.~\ref{fig:smeansdevs_vs_absx_diffw}.  With increasing $N_e$, $\Delta
\sigma_{zz}/\sigma_{zz}$ seems to saturate to a well-defined limit (for a given
$w$).  The relative stress fluctuations at fixed $x$ seem to approximately
satisfy: $\Delta \sigma_{zz}/\sigma_{zz} \propto w^{-2/3}$.  While an
explanation of this possible ``scaling'' is still lacking, we believe it is quite
surprising that the fluctuations at small values of $w$ (clearly related to the
force and particle position fluctuations) share the same (approximate) scaling
with those at large values of $w$, which sample large-scale fluctuations and
the spatial variation of the stress.

The dependence of the stress field (for $w$  within the plateau) and 
its
fluctuations on the absolute horizontal distance from the load, $|x|$, is
presented in Fig.~\ref{fig:smeansdevs_vs_absx_diffw}.  The fluctuations
($\Delta \sigma$) seem to decay nearly exponentially with $x$, at least not too
far from $x=0$.  The decay length depends on $w$. However, the mean stress
decays faster than exponentially at large $x$ (basically as a Gaussian), in
conformity with the linear elastic solution for this case; the sharp drop near
$|x|=50\left<d\right>$ is due to the fact that the response becomes negative
over a small interval, as expected from linear
elasticity~\cite{Serero01,Goldenberg02}. A similar study of a Lennard-Jones
glass in the linear response regime, with a localized force applied in the
interior (rather than the boundary) is presented in Fig.~7
of~\cite{Leonforte04} for 2D and Fig.~8 of~\cite{Leonforte05} for 3D systems
(the dependence on $w$ is not discussed in~\cite{Leonforte04,Leonforte05}).  In
our case the mean stress (calculated at the boundary) decays faster than
exponentially, while in~\cite{Leonforte04,Leonforte05} the stress is calculated
in the bulk, and decays algebraically (both results are consistent with linear
elasticity). This renders the relative fluctuations, in our case,
non-negligible at large distances, as opposed to the decay with distance found
in~\cite{Leonforte04,Leonforte05}. Our findings suggest that this decay only
characterizes intermediate distances (or possibly a wall effect), whereas
asymptotically the relative standard deviation is finite (see
Fig.~\ref{fig:smeansdevs_vs_absx_diffw}).

In summary, our findings indicate that the magnitudes of the local gradients of
the macroscopic fields (stress in the above case) determine the widths of the
plateaus. Although large gradients are expected in nonuniformly forced small
systems, the size of the system is not the main factor that limits these
widths.  The mere existence of the plateaus (which can be as small as 3-5
particle diameters) suggests that continuum theories may be valid for granular
and mesoscopic solid systems, but one may need to go beyond simple linear
descriptions. The saturation of the results for small ensembles (e.g., 40-50
realizations) due to the short range of force correlations suggests that
appropriate constitutive relations can be derived for such systems. Although
this Letter is restricted to the linear response regime, we believe that the
plateaus will continue to exist even near fluidization (they do exist in the
fluid regime), hence, the above approach should be relevant to a rather large
range of loads.

\begin{acknowledgments}
  We thank E.~Cl{\'e}ment, F.~L{\'e}onforte, A.~Tanguy and \mbox{J.-L.~Barrat}
  for useful discussions. C.~G.\ acknowledges support from a Chateaubriand
  grant. A.~P.~F.~A is grateful for the support of the CNPq and CAPES (Brazil).
  A.~P.~F.~A, C.~G., I.~G., and P.~C.\ express their gratitude to
  Arc-en-Ciel--Keshet exchange program.  I.~G.\ gratefully acknowledges partial
  support from the ISF, grant no.\ 689/04, GIF, grant no.\ 795/2003 and BSF,
  \mbox{grant no.\ 2004391}.
\end{acknowledgments}

\end{document}